\documentstyle[amssymb,11pt]{article}

\newlength{\minitwocolumn}
\font\teneufm=eufm10
\font\seveneufm=eufm7
\font\fiveeufm=eufm5
\newfam\eufmfam
\textfont\eufmfam=\teneufm
\scriptfont\eufmfam=\seveneufm
\scriptscriptfont\eufmfam=\fiveeufm

\makeatletter
\@addtoreset{equation}{section}
\makeatother

\newtheorem{thm}{Theorem}[section]
\newtheorem{prop}[thm]{Proposition}

\newtheorem{cor}[thm]{Corollary}

\title{\bf
\huge{\bf
Diagonalization of
boundary transfer matrix\\
for the $U_{q,p}(\widehat{\sl sl}(3,{\mathbb C}))$ ABF model}
}
\begin{document}
\maketitle
\begin{center}
{Takeo KOJIMA}
\\~\\
{\it
Department of Mathematics,
College of Science and Technology,
Nihon University,\\
Surugadai, Chiyoda-ku, Tokyo 101-0062, 
JAPAN}
\end{center}
~\\
\begin{abstract}
We construct a free field realization of the ground state
of the boundary transfer matrix for 
the $U_{q,p}(\widehat{sl}(3,{\mathbb C}))$ ABF model.
Using this ground state and type-II vertex operator,
we have a diagonalization of the boundary
transfer matrix.
\end{abstract}

~\\

\section{Introduction}

The vertex operator approach 
\cite{DFJMN, JM}
provides a powerful method to study solvable models.
In \cite{DFJMN} authors diagonalized 
the XXZ-transfer matrix on infinite spin chain
by vertex operators. 
Sklyanin \cite{Sklyanin}
begin systematic approach to 
boundary condition generalization in the framework
of the algebraic Bethe ansatz.
In periodic boundary condition,
we construct a family of 
commuting transfer matrix from a solution of the
Yang-Baxter equation.
Sklyanin showed
that similar construction is possible with the aid
of a solution of the boundary Yang-Baxter equation.
In open boundary condition,
we construct a family of commuting transfer matrix from both a solution of the
Yang-Baxter equation and a solution of the boundary Yang-Baxter
equation.
In \cite{JKKKM} the authors diagonalized 
the boundary XXZ-transfer matrix on 
semi-infinite spin chain
by vertex operators. 
The vertex operator approach was extended to
higher-rank boundary 
XXZ-model in \cite{FK}

The vertex operator approach to solvable model
was originally formulated for vertex type model such as 
the XXZ-model \cite{DFJMN}, and then extended to 
the face type model
such as the Andrews-Baxter-Forrester (ABF) model
\cite{ABF, LP}. 
The ABF model is described by the elliptic algebra 
$U_{q,p}(\widehat{sl}(2,{\mathbb C}))$.
The $U_{q,p}(\widehat{sl}(3,{\mathbb C}))$ ABF model
we are going to study in this paper
were introduced in \cite{JMO}, as the higher-rank 
generlization of the ABF model introduced
in \cite{ABF}.
In this paper we diagonalize
the boundary transfer matrix
for the
$U_{q,p}(\widehat{sl}(3,{\mathbb C}))$ ABF model
by vertex operator approach.
We construct a free field realization of
the ground-state of the commuting
boundary transfer matrix.
Using this groundstate and type-II vertex operators 
for the elliptic algebra,
we get a diagonalization
of the boundary transfer matrix of
the $U_{q,p}(\widehat{sl}(3,{\mathbb C}))$ ABF model.
The result of this paper gives a higher-rank
generalization of the ABF model \cite{MW}.

The text is organized as follows.
In section 2 we recall
the boundary $U_{q,p}(\widehat{sl}(3,{\mathbb C}))$ ABF model,
and introduce the boundary transfer matrix.
In section 3
we recall the free field realization of the vertex operators.
In section 4 we construct a free field realization of
the ground-state of the boundary transfer matrix,
and give a diagonalization
of the boundary transfer matrix by
using the type-II vertex opeartors.

\section{Boundary $U_{q,p}(\widehat{sl}(3,{\mathbb C}))$ 
ABF model}

We recall the boundary 
$U_{q,p}(\widehat{sl}(3,{\mathbb C}))$ ABF model.

\subsection{Bulk Boltzmann weights}

The $U_{q,p}(\widehat{sl}(3,{\mathbb C}))$ ABF model
has two parameters $x$ and $r$.
We assume $0<x<1$ and $r \geqq 5~(r \in {\mathbb Z})$.
We set the elliptic theta function $[u]$ by
\begin{eqnarray}
~[u]=x^{\frac{u^2}{r}-u}\Theta_{x^{2r}}(x^{2u}).
\end{eqnarray}
Here we have used
\begin{eqnarray}
\Theta_q(z)&=&(q,q)_\infty (z;q)_\infty (q/z;q)_\infty,\\
(z;q_1,q_2,\cdots,q_m)_\infty&=&\prod_{j_1,j_2,\cdots,j_m=0}^\infty
(1-q_1^{j_1}q_2^{j_2}\cdots q_m^{j_m}z).
\end{eqnarray}
Let $\epsilon_\mu (1\leqq \mu \leqq 3)$ be the orthonormal
basis of ${\mathbb R}^3$ with the inner
product $(\epsilon_\mu |\epsilon_\nu)=\delta_{\mu,\nu}$.
Let us set $\bar{\epsilon}_\mu=\epsilon_\mu-\epsilon$
where $\epsilon=\frac{1}{3}\sum_{\nu=1}^3 \epsilon_\nu$.
The type $\widehat{sl}(3,{\mathbb C})$ weight lattice is the
linear span of $\bar{\epsilon}_\mu$.
\begin{eqnarray}
P=\sum_{\mu=1}^3 {\mathbb Z}\bar{\epsilon}_\mu.
\end{eqnarray}
Note $\sum_{\mu=1}^3 \bar{\epsilon}_\mu=0$.
Let the simple root $\alpha_\mu=
\bar{\epsilon}_\mu-\bar{\epsilon}_{\mu+1}$ $(\mu=1,2)$.
For $a \in P$ we set
\begin{eqnarray}
a_{\mu,\nu}=a_\mu-a_\nu,~~~
a_\mu=(a+\rho|\bar{\epsilon}_\mu),
\end{eqnarray}
where $\rho=2 \bar{\epsilon}_1+\bar{\epsilon}_2$.
The Boltzmann weights 
$
W\left(\left.\begin{array}{cc}
a&b\\
c&d
\end{array}\right|u\right)$
are given by
\begin{eqnarray}
&&W\left(\left.
\begin{array}{cc}
a+2\bar{\epsilon}_\mu & a+\bar{\epsilon}_\mu\\
a+\bar{\epsilon}_\mu & a
\end{array}\right|u\right)=r_1(u),\\
&&W\left(\left.
\begin{array}{cc}
a+\bar{\epsilon}_\mu+\bar{\epsilon}_\nu & 
a+\bar{\epsilon}_\mu\\
a+\bar{\epsilon}_\nu & a
\end{array}\right|u\right)=r_1(u)\frac{[u][a_{\mu,\nu}-1]}
{[u-1][a_{\mu,\nu}]},\\
&&W\left(\left.
\begin{array}{cc}
a+\bar{\epsilon}_\mu+\bar{\epsilon}_\nu 
& a+\bar{\epsilon}_\nu\\
a+\bar{\epsilon}_\nu & a
\end{array}\right|u\right)=r_1(u)\frac{[u-a_{\mu,\nu}][1]}{
[u-1][a_{\mu,\nu}]}.
\end{eqnarray}
Otherwise are zero.
The function $r_1(u)$ is given by 
\begin{eqnarray}
r_1(u)&=&z^{\frac{r-1}{r}\frac{2}{3}}\frac{h_1(z^{-1})}{h(z)},~~~
h_1(z)=\frac{
(x^2z;x^{2r},x^6)_\infty 
(x^{2r+4}z;x^{2r},x^6)_\infty}{
(x^{2r}z;x^{2r},x^6)_\infty 
(x^6z;x^{2r},x^6)_\infty}.
\end{eqnarray}
Here we have used $z=x^{2u}$.
The Boltzmann weights satisfy the following relations.
\\
(1)Yang-Baxter equation :
\begin{eqnarray}
&&\sum_{g}
W\left(\left.\begin{array}{cc}
d&e\\
c&g
\end{array}
\right|u_1\right)
W\left(\left.\begin{array}{cc}
c&g\\
b&a
\end{array}
\right|u_2\right)
W\left(\left.\begin{array}{cc}
e&f\\
g&a
\end{array}
\right|u_1-u_2\right)
\nonumber\\
&=&
\sum_{g}
W\left(\left.\begin{array}{cc}
g&f\\
b&a
\end{array}
\right|u_1\right)
W\left(\left.\begin{array}{cc}
d&e\\
g&f
\end{array}
\right|u_2\right)
W\left(\left.\begin{array}{cc}
d&g\\
c&b
\end{array}
\right|u_1-u_2\right).
\end{eqnarray}
(2)The first inversion relation :
\begin{eqnarray}
\sum_{g}
W\left(\left.\begin{array}{cc}
c&g\\
b&a
\end{array}
\right|-u\right)
W\left(\left.\begin{array}{cc}
c&d\\
g&a
\end{array}
\right|u\right)
=\delta_{b,d}.
\end{eqnarray}
(3)The second inversion relation :
\begin{eqnarray}
\sum_{g}G_g
W\left(\left.\begin{array}{cc}
g&b\\
d&c
\end{array}
\right|3-u\right)
W\left(\left.\begin{array}{cc}
g&d\\
b&a
\end{array}
\right|u\right)
=\delta_{a,c}\frac{G_{b}G_d}{G_a}.
\end{eqnarray}
where $G_a=[a_{1,2}][a_{1,3}][a_{2,3}]$.
The Boltzmann weights are related to
the elliptic algebra $U_{q,p}(\widehat{sl}(3,{\mathbb C}))$.

\subsection{Boundary Boltzmann weights}

In \cite{BFKZ} the boundary Boltzmann weights 
$
K\left(
\left.\begin{array}{cc}
&b\\
a&\\
&c
\end{array}
\right|u\right)$
are given by
\begin{eqnarray}
K\left(
\left.\begin{array}{cc}
&k\\
k+\bar{\epsilon}_\mu&\\
&k
\end{array}
\right|u\right)=
z^{\frac{2}{r}(-\frac{r-1}{3}+(\bar{\epsilon}_1|k))}
\frac{h(z)}{h(z^{-1})}
\frac{[c-u][k_{1,\mu}+c+u]}
{[c+u][k_{1,\mu}+c-u]},~(c \in {\mathbb R}; \mu=1,2,3).
\end{eqnarray}
Otherwise are zero.
The function $h(z)$ is given in (\ref{def:h}).
They satisfy the boundary Yang-Baxter equation,
\begin{eqnarray}
&&\sum_{f,g}
W\left(\left.\begin{array}{cc}
c&f\\
b&a
\end{array}\right|u-v\right)
W\left(\left.\begin{array}{cc}
c&d\\
f&g
\end{array}
\right|u+v\right)
K\left(\left.\begin{array}{cc}
~&g\\
f&\\
~&a
\end{array}
\right|u\right)
K\left(\left.\begin{array}{cc}
~&e\\
d&\\
~&g
\end{array}
\right|v\right)
\nonumber\\
&=&
\sum_{f,g}
W\left(\left.\begin{array}{cc}
c&d\\
f&e
\end{array}\right|u-v\right)
W\left(\left.\begin{array}{cc}
c&f\\
b&g
\end{array}
\right|u+v\right)
K\left(\left.\begin{array}{cc}
~&e\\
f&\\
~&g
\end{array}
\right|u\right)
K\left(\left.\begin{array}{cc}
~&g\\
b&\\
~&a
\end{array}
\right|v\right).
\end{eqnarray}

\subsection{Vertex operator}

Following the general scheme of algebraic approach
in solvable lattice models,
we give the type-I vertex operators.
Let us consider the corner transfer matrices
$A(z),B(z),C(z),D(z)$ which represent
NW, SW, SE, NE quadrants, respectively.
The space ${\cal H}_{l,k}$ of 
the eigenvectors of $A(z)$ is parametrized
by $l,k \in P$.
Let us introduce the type-I vertex operator
$\Phi^{(a,b)}(z)$.
We denote by 
$\Phi^{(k+\bar{\epsilon}_j,k)}(z^{-1})$
the half-infinite transfer matrix
extending to infinity in the north.
This is an operator
\begin{eqnarray}
\Phi^{(k+\bar{\epsilon}_j,k)}(z^{-1}) :
{\cal H}_{l,k} \rightarrow {\cal H}_{l,k+\bar{\epsilon}_j}.
\nonumber
\end{eqnarray}
The opeartor $\Phi^{(a,b)}(z)=0$ for $a-b\neq \bar{\epsilon}_j$.
Similarily, we introduce dual type-I vertex operator
$\Phi^{*(a,b)}(z)$.
We denote by
$\Phi^{*(k,k+\bar{\epsilon}_j)}(z)$
the half-infinite transfer matrix extending to infinity
in the west.
This is an operator
\begin{eqnarray}
\Phi^{*(k,k+\bar{\epsilon}_j)}(z) :
{\cal H}_{l,k+\bar{\epsilon}_j} \rightarrow 
{\cal H}_{l,k}.\nonumber
\end{eqnarray}
The opeartor $\Phi^{(a,b)}(z)=0$ for $b-a\neq \bar{\epsilon}_j$.
They satisfy the following relations.\\
(1)Commutation relation
\begin{eqnarray}
\Phi^{(c,b)}(z_1)
\Phi^{(b,a)}(z_2)
=\sum_{d}
W\left(\left.\begin{array}{cc}
c&d\\
b&a
\end{array}
\right|u_1-u_2\right)
\Phi^{(c,d)}(z_2)
\Phi^{(d,a)}(z_1).
\end{eqnarray}
(2)Inversion relation
\begin{eqnarray}
\sum_{g}\Phi^{*(a,g)}(z)\Phi^{(g,a)}(z)=1,~~~
\Phi^{(a,b)}(z)\Phi^{*(b,c)}(z)=\delta_{a,c}.
\end{eqnarray}
Later we give a free field realization of
the vertex operators acting on the bosonic Fock space.

\subsection{Boundary transfer matrix}

We define the boundary transfer matrix
\begin{eqnarray}
T_B^{(k)}(z)=\sum_{j=1,2,3}
\Phi^{*(k,k+\bar{\epsilon}_j)}(z)
K\left(\left.
\begin{array}{cc}
~& k\\
k+\bar{\epsilon}_j&\\
~& k
\end{array}
\right|u\right)
\Phi^{(k+\bar{\epsilon}_j,k)}(z^{-1}).
\end{eqnarray}
The boundary Yang-Baxter equation implies
the commutativity.
\begin{eqnarray}
~[T_B^{(k)}(z_1),T_B^{(k)}(z_2)]=0.
\end{eqnarray}
Our problem of this paper is to diagonalize 
the boundary transfer matrix
$T_B^{(k)}(z)$.

\section{Free field realization}

We give a free field realization of the vertex operators
\cite{AJMP, FKQ}.

\subsection{Boson}

We set the bosonic oscillators $\beta_m^i, (i=1,2;
m \in {\mathbb Z})$ by
\begin{eqnarray}
~[\beta_m^j,\beta_n^k]=
\left\{
\begin{array}{cc}
\displaystyle
m \frac{[(r-1)m]_x }{[rm]_x}
\frac{[2m]_x}{[3m]_x}\delta_{m+n,0}
&(j=k)\\
\displaystyle
-m x^{3m~sgn(j-k)}
\frac{[(r-1)m]_x}{[rm]_x}
\frac{[m]_x}{[3m]_x}\delta_{m+n,0}
&(j \neq k).
\end{array}
\right.
\end{eqnarray}
Here the symbol $[a]_x=\frac{x^a-x^{-a}}{x-x^{-1}}$.
Let us set $\beta_m^3$ by
$\sum_{j=1}^3x^{-2jm}\beta_m^j=0$.
The above commutation relations
are valid for all $1\leqq j,k \leqq 3$.
We also introduce
the zero-mode operators $P_\alpha, Q_\alpha$, $(\alpha \in P)$
by
\begin{eqnarray}
~[iP_\alpha,Q_\beta]=(\alpha|\beta),~~(\alpha,\beta \in P).
\end{eqnarray}
In what follows we 
deal with the bosonic Fock space ${\cal F}_{l,k} (l,k\in P)$
generated by $\beta_{-m}^j (m>0)$ over the vacuum vector 
$|l,k \rangle$ :
\begin{eqnarray}
{\cal F}_{l,k}={\mathbb C}[\{\beta_{-1}^j,
\beta_{-2}^j,\cdots\}_{j=1,2,3}]|l,k\rangle,
\end{eqnarray}
where
\begin{eqnarray}
&&\beta_m^j |k,l\rangle=0,~~(m>0),\nonumber
\\
&&P_\alpha |l,k\rangle=\left(\alpha\left|
\sqrt{\frac{r}{r-1}}l-\sqrt{\frac{r-1}{r}}k
\right.\right)|l,k\rangle,\nonumber\\
&&|l,k\rangle=e^{i\sqrt{\frac{r}{r-1}}Q_l-
\sqrt{\frac{r-1}{r}}Q_k}|0,0\rangle.\nonumber
\end{eqnarray}

\subsection{Vertex operator}

We give a free field realization of
the type-I vertex operators \cite{AJMP},
associated with the elliptic algebra
$U_{q,p}(\widehat{sl}(3,{\mathbb C}))$
\cite{JKOS, KK}.
Let us set $P(z),Q(z),R_-^j(z),S_-^j(z) (j=1,2)$ by
\begin{eqnarray}
&&P(z)=\sum_{m>0}\frac{1}{m}\beta_{-m}^1z^{m},~~
Q(z)=-\sum_{m>0}\frac{1}{m}\beta_m^1z^{-m},\\
&&
R_-^j(z)=
-\sum_{m>0}\frac{1}{m}(\beta_{-m}^j-\beta_{-m}^{j+1})x^{jm}z^m,~~
S_-^j(z)=
\sum_{m>0}\frac{1}{m}(\beta_m^j-\beta_m^{j+1})x^{-jm}z^{-m}.
\end{eqnarray}
Let us set the basic operators $U(z), F_{\alpha_1}(z),
F_{\alpha_2}(z)$ on the Fock space ${\cal F}_{l,k}$.
\begin{eqnarray}
U(z)&=&
z^{\frac{r-1}{3r}}
e^{-i\sqrt{\frac{r-1}{r}}Q_{\bar{\epsilon}_1}}
z^{-\sqrt{\frac{r-1}{r}}P_{\bar{\epsilon}_1}}
e^{P(z)}e^{Q(z)},\\
F_{\alpha_j}(z)&=&
z^{\frac{r-1}{r}}
e^{i\sqrt{\frac{r-1}{r}}Q_{\alpha_j}}
z^{\sqrt{\frac{r-1}{r}}P_{\alpha_j}}
e^{R_-^j(z)}e^{S_-^j(z)}.
\end{eqnarray}
In what follows we set
\begin{eqnarray}
\pi_\mu=\sqrt{r(r-1)}P_{\bar{\epsilon}_\mu},~~
\pi_{\mu \nu}=\pi_\mu-\pi_\nu.
\end{eqnarray}
Then $\pi_{\mu \nu}$ acts on ${\cal F}_{l,k}$
as an integer $(\epsilon_\mu-\epsilon_\nu|rl-(r-1)k)$.
We give the free field realization of
the type-I vertex operators.
\begin{eqnarray}
\Phi_1(z)&=&U(z),\\
\Phi_2(z)&=&\oint_{C_1} \frac{dw_1}{w_1} U(z)
F_{\alpha_1}(w_1)\frac{[v_1-u+\frac{1}{2}-\pi_{1,2}]}
{[v_1-u-\frac{1}{2}]},\\
\Phi_3(z)&=&\oint \oint_{C_2} \frac{dw_1}{w_1}
\frac{dw_2}{w_2} U(z)
F_{\alpha_1}(w_1)F_{\alpha_2}(w_2)
\frac{[v_1-u+\frac{1}{2}-\pi_{1,3}]}
{[v_1-u-\frac{1}{2}]}
\frac{[v_2-v_1+\frac{1}{2}-\pi_{2,3}]}
{[v_2-v_1-\frac{1}{2}]}.\nonumber\\
\end{eqnarray}
Here we set $z=x^{2u}, w_j=x^{2v_j} (j=1,2)$.
We take the integration contours to be simple closed 
curves around the origin satisfying
\begin{eqnarray}
&&C_1: x|z|<|w_1|<x^{-1}|z|,\nonumber\\
&&C_2: x|z|<|w_1|<x^{-1}|z|,~~
x|w_1|<|w_2|<x^{-1}|w_1|.
\nonumber
\end{eqnarray}
We identify
$\Phi^{(k+\bar{\epsilon}_j,k)}(z)=\Phi_j(z)$.

Let us introduce the type-II vertex operators
\cite{FKQ}, associated with the elliptic algebra
$U_{q,p}(\widehat{sl}(3,{\mathbb C}))$.
The type-II vertex operators
represents the excitations.
Let us set $r^*=r-1$.
Let us set the elliptic theta function $[u]^*$ by
\begin{eqnarray}
~[u]^*=x^{\frac{u^2}{r^*}-u}\Theta_{x^{2r^*}}(x^{2u}).
\end{eqnarray}
Let us set $P^*(z),Q^*(z),R_+^j(z),S_+^j(z) (j=1,2)$ by
\begin{eqnarray}
&&
P^*(z)=-\sum_{m>0}\frac{[rm]_x}{m[r^*m]_x}\beta_{-m}^1z^{m},~~
Q^*(z)=\sum_{m>0}\frac{[rm]_x}{m[r^*m]_x}\beta_m^1z^{-m},\\
&&
R_+^j(z)=
\sum_{m>0}\frac{[rm]_x}{m[r^*m]_x}
(\beta_{-m}^j-\beta_{-m}^{j+1})x^{jm}z^m,~~
S_+^j(z)=
-\sum_{m>0}\frac{[rm]_x}{m[r^*m]_x}
(\beta_m^j-\beta_m^{j+1})x^{-jm}z^{-m}.\nonumber\\
\end{eqnarray}
Let us set the basic operators $V(z), E_{\alpha_1}(z),
E_{\alpha_2}(z)$ on the Fock space ${\cal F}_{l,k}$.
\begin{eqnarray}
V(z)&=&
z^{\frac{r}{3r^*}}
e^{i\sqrt{\frac{r}{r^*}}Q_{\bar{\epsilon}_1}}
z^{\sqrt{\frac{r}{r^*}}P_{\bar{\epsilon}_1}}
e^{P^*(z)}e^{Q^*(z)},\\
F_{\alpha_j}(z)&=&
z^{\frac{r}{r^*}}
e^{-i\sqrt{\frac{r}{r^*}}Q_{\alpha_j}}
z^{-\sqrt{\frac{r}{r^*}}P_{\alpha_j}}
e^{R_+^j(z)}e^{S_+^j(z)}.
\end{eqnarray}
We give a free field realization of
the type-II vertex operators.
\begin{eqnarray}
\Psi_1^*(z)&=&V(z),\\
\Psi_2^*(z)&=&\oint_{\tilde{C}_1} \frac{dw_1}{w_1} V(z)
E_{\alpha_1}(w_1)\frac{[v_1-u-\frac{1}{2}+\pi_{1,2}]^*}
{[v_1-u+\frac{1}{2}]^*},\\
\Psi_3^*(z)&=&\oint \oint_{\tilde{C}_2} \frac{dw_1}{w_1}
\frac{dw_2}{w_2} V(z)
E_{\alpha_1}(w_1)E_{\alpha_2}(w_2)
\frac{[v_1-u-\frac{1}{2}+\pi_{1,3}]^*}
{[v_1-u+\frac{1}{2}]^*}
\frac{[v_2-v_1-\frac{1}{2}+\pi_{2,3}]^*}
{[v_2-v_1+\frac{1}{2}]^*}.\nonumber\\
\end{eqnarray}
The integration contours $\tilde{C}_1$
encloses the poles at $w_1=x^{-1+2sr^*}z, (s=0,1,2,\cdots)$,
but not the poles at $w_1=x^{1-2sr^*}z, (s=0,1,2,\cdots)$.
The integration of $\Psi_3^*(z)$ is carried out
in the order $w_2, w_1$
along the contours $\tilde{C}_2$
which encloses the poles
the poles at $w_{j+1}
=x^{-1+2sr^*}w_j, (s=0,1,2,\cdots; j=1,2)$,
but not the poles at 
$w_{j+1}=x^{1-2sr^*}w_j, (s=0,1,2,\cdots; j=1,2)$.
Here we set $w_0=z$.
Let us set
the type-II Boltzmann weights 
$
W^*\left(\left.\begin{array}{cc}
a&b\\
c&d
\end{array}\right|u\right)$
are given by
\begin{eqnarray}
&&W^*\left(\left.
\begin{array}{cc}
a+2\bar{\epsilon}_\mu & a+\bar{\epsilon}_\mu\\
a+\bar{\epsilon}_\mu & a
\end{array}\right|u\right)=r_1^*(u),\\
&&W^*\left(\left.
\begin{array}{cc}
a+\bar{\epsilon}_\mu+\bar{\epsilon}_\nu & 
a+\bar{\epsilon}_\mu\\
a+\bar{\epsilon}_\nu & a
\end{array}\right|u\right)=r_1^*(u)
\frac{[u]^*[a_{\mu,\nu}-1]^*}
{[u-1]^*[a_{\mu,\nu}]^*},\\
&&W^*\left(\left.
\begin{array}{cc}
a+\bar{\epsilon}_\mu+\bar{\epsilon}_\nu 
& a+\bar{\epsilon}_\nu\\
a+\bar{\epsilon}_\nu & a
\end{array}\right|u\right)=r_1^*(u)
\frac{[u-a_{\mu,\nu}]^*[1]^*}{
[u-1]^*[a_{\mu,\nu}]^*}.
\end{eqnarray}
Otherwise are zero.
The function $r_1^*(u)$ is given by 
\begin{eqnarray}
r_1^*(u)&=&z^{\frac{r}{r^*}\frac{2}{3}}\frac{h_1^*(z^{-1})}{h^*(z)},~~~
h_1^*(z)=\frac{
(z;x^{2r^*},x^6)_\infty 
(x^{4}z;x^{2r^*},x^6)_\infty}{
(x^{2r^*}z;x^{2r^*},x^6)_\infty 
(x^6z;x^{2r^*},x^6)_\infty}.
\end{eqnarray}
The Boltzmann weights satisfy the following relations.
\\
(1)Yang-Baxter equation :
\begin{eqnarray}
&&\sum_{g}
W^*\left(\left.\begin{array}{cc}
d&e\\
c&g
\end{array}
\right|u_1\right)
W^*\left(\left.\begin{array}{cc}
c&g\\
b&a
\end{array}
\right|u_2\right)
W^*\left(\left.\begin{array}{cc}
e&f\\
g&a
\end{array}
\right|u_1-u_2\right)
\nonumber\\
&=&
\sum_{g}
W^*\left(\left.\begin{array}{cc}
g&f\\
b&a
\end{array}
\right|u_1\right)
W^*\left(\left.\begin{array}{cc}
d&e\\
g&f
\end{array}
\right|u_2\right)
W^*\left(\left.\begin{array}{cc}
d&g\\
c&b
\end{array}
\right|u_1-u_2\right).
\end{eqnarray}
(2)The first inversion relation :
\begin{eqnarray}
\sum_{g}
W^*\left(\left.\begin{array}{cc}
c&g\\
b&a
\end{array}
\right|-u\right)
W^*\left(\left.\begin{array}{cc}
c&d\\
g&a
\end{array}
\right|u\right)
=\delta_{b,d}.
\end{eqnarray}
(3)The second inversion relation :
\begin{eqnarray}
\sum_{g}G_g^*
W^*\left(\left.\begin{array}{cc}
g&b\\
d&c
\end{array}
\right|3-u\right)
W^*\left(\left.\begin{array}{cc}
g&d\\
b&a
\end{array}
\right|u\right)
=\delta_{a,c}\frac{G_{b}^*G_d^*}{G_a^*}.
\end{eqnarray}
where $G_a^*=[a_{1,2}]^*[a_{1,3}]^*[a_{2,3}]^*$.
The type-II vertex operators satisfy
the following relations.
\begin{eqnarray}
\Phi^*_{\mu_1}(z_1)
\Phi^*_{\mu_2}(z_2)
=\sum_{\epsilon_{\mu_1'}+{\epsilon}_{\mu_2'}=
\epsilon_{\mu_1}+{\epsilon}_{\mu_2}
}
W^*\left(\left.\begin{array}{cc}
k+\bar{\epsilon}_{\mu_1}+\bar{\epsilon}_{\mu_2}&
k+\bar{\epsilon}_{\mu_1'}\\
k+\bar{\epsilon}_{\mu_2}&k
\end{array}
\right|u_2-u_1\right)
\Phi^*_{\mu_2'}(z_2)
\Phi^*_{\mu_1'}(z_1).\nonumber\\
\end{eqnarray}
The type-I and type-II vertex operators commute
modulo the "energy function" $\chi(z)$.
\begin{eqnarray}
\Phi_{\mu_1}(z_1)\Psi^*_{\mu_2}(z_2)=
\chi(z_1/z_2)\Psi^*_{\mu_2}(z_2)
\Phi_{\mu_1}(z_1).
\end{eqnarray}
Here we have set
\begin{eqnarray}
\chi(z)=z^{\frac{2}{3}}
\frac{(x^5z^{-1};x^6)_\infty (xz;x^6)_\infty}{
(xz^{-1};x^6)_\infty (x^{5}z;x^6)_\infty}.\label{def:chi}
\end{eqnarray}

\section{Boundary state}

In this section we construct the free field realization
of the ground state of the boundary transfer matrix of 
the $U_{q,p}(\widehat{sl}(3,{\mathbb C}))$ ABF model, 
which give higher-rank generalization of \cite{MW}.
We construct the free field realization of
the ground state on the Fock space ${\cal F}_{l,k}$.
Precisely the Fock space ${\cal F}_{l,k}$ is different from
the space of the state ${\cal H}_{l,k}$.
The space of state ${\cal H}_{l,k}$ is obtained
by the cohomological argument from the Fock space 
${\cal F}_{l,k}$ by using
the so-called screening operators.
In this paper we omit the detailed 
of the screening operators and
consider the operators acting on 
the Fock space ${\cal F}_{l,k}$.

\subsection{Bogoliubov transformation}

The commutation relations of bosons $\beta_m^j$ are not
symmetric. Hence
it is convenient to introduce
new generators of bosons $\alpha_m^1, \alpha_m^2$
whose commutation relations are symmetric.
\begin{eqnarray}
\alpha_m^1=x^{-m}(\beta_m^1-\beta_m^2),~~
\alpha_m^2=x^{-2m}(\beta_m^2-\beta_m^3).
\end{eqnarray}
They satisfy the following commutation relations.
\begin{eqnarray}
~[\alpha_m^j,\alpha_n^k]=
\left\{
\begin{array}{cc}
\displaystyle
m \frac{[(r-1)m]_x}{[rm]_x}\frac{[2m]_x}{[m]_x}\delta_{m+n,0}
&(j=k)\\
\displaystyle
-m \frac{[(r-1)m]_x}{[rm]_x}\delta_{m+n,0}
&(j\neq k).
\end{array}
\right.
\end{eqnarray}
Let us set
\begin{eqnarray}
F_0=-\sum_{m>0}\frac{1}{m}\frac{[rm]_x}{[(r-1)m]_x[3m]_x}
([2m]_x\alpha_{-m}^1 \alpha_{-m}^1
+2[m]_x \alpha_{-m}^1 \alpha_{-m}^2
+[2m]_x \alpha_{-m}^2 \alpha_{-m}^2).
\label{def:f0}
\end{eqnarray}
The adjioint action of $e^{F_0}$ has the effect of 
a Bogoliubov transformation,
\begin{eqnarray}
&&
e^{-F_0}\alpha_m^j e^{F_0}=
\alpha_m^j-\alpha_{-m}^j~~(m>0, j=1,2),\\
&&
e^{-F_0}\alpha_{-m}^j e^{F_0}
=\alpha_{-m}^j~~(m>0, j=1,2),
\end{eqnarray}
and
\begin{eqnarray}
&&
e^{-F_0}\beta_m^1 e^{F_0}=-\beta_{-m}^1~~(m>0),\\
&&
e^{-F_0}\beta_m^2 e^{F_0}=(x^{2m}-1)\beta_{-m}^1-
x^{2m}\beta_{-m}^2~~(m>0),
\\
&&
e^{-F_0}\beta_{-m}^j e^{F_0}
=\beta_{-m}^j~~(m>0, j=1,2).
\end{eqnarray}

\begin{prop}~
\begin{eqnarray}
&&e^{-F_0}e^{Q(z)}e^{F_0}= 
\sqrt{\frac
{
(x^{2r+4}z^{-2};x^{2r},x^6)_\infty
(x^2z^{-2};x^{2r},x^6)_\infty
}
{
(x^{2r}z^{-2};x^{2r},x^6)_\infty 
(x^6z^{-2};x^{2r},x^6)_\infty}}
e^{P(1/z)}e^{Q(z)},\\
&&e^{-F_0}e^{S_-^j(z)}e^{F_0}=
\sqrt{\frac{
(1-z^{-2})(x^{2r-2}z^{-2};x^{2r})_\infty
}{
(x^{2}z^{-2};x^{2r})_\infty}}
e^{R_-^j(1/z)}e^{S_-^j(z)}~~(j=1,2).
\end{eqnarray}
\end{prop}

\subsection{Boundary state}

Let us set the bosonic operator $F$ by
\begin{eqnarray}
F=F_0+F_1,
\end{eqnarray}
where $F_0$ is given in (\ref{def:f0})
and $F_1$ is given by
\begin{eqnarray}
F_1=\sum_{m>0}
(D_1(m)\beta_{-m}^1+D_2(m)\beta_{-m}^2).
\end{eqnarray}
Here we set
\begin{eqnarray}
D_1(m)&=&\frac{x^{-m}[(r-2c+2-2\pi_{1,3})m]_x
-[(r-2c-1)m]_x+x^{(r-2c-2\pi_{1,2})m}[m]_x
}{m[(r-1)m]_x}\nonumber\\
&-&\theta_m\left(\frac{x^{-\frac{m}{2}}
[m]_x[rm/2]_x^+}{
m[(r-1)m/2]_x}\right),\\
D_2(m)&=&\frac{
x^m([(r-2c+2-2\pi_{1,3})m]_x-[(r-2c-2\pi_{1,2})m]_x)
}{m[(r-1)m]_x}
-\theta_m \left(\frac{x^m[rm/2]_x^+[m/2]_x}{m
[(r-1)m/2]_x}\right),
\nonumber\\
\end{eqnarray}
where 
\begin{eqnarray}
~[a]_x^+=a^x+a^{-x},~~~\theta_m(x)=\left\{
\begin{array}{cc}
x & (m:even)\\
0 & (m:odd)
\end{array}\right..\nonumber
\end{eqnarray}
Let us set the vector
\begin{eqnarray}
|B\rangle_{l,k}=e^{F}|l,k \rangle.
\end{eqnarray}
We call this vector $|B\rangle_{l,k}$ the boundary state.

\begin{prop}~~The boundary state $|B\rangle_{l,k}$
has the following properties.
\begin{eqnarray}
e^{Q(z)}|B\rangle_{l.k}&=&h(z)e^{P(1/z)}|B\rangle_{l,k},\\
e^{S_-^j(w)}|B\rangle_{l,k}&=&g_j(w)e^{R_-^j(1/w)}
|B\rangle_{l,k},~(j=1,2).
\end{eqnarray}
Here the fuctions $h(z), g_j(w) (j=1,2)$ are given by
\begin{eqnarray}
h(z)&=&
\frac{(x^{2r+4}z^{-2};x^{2r},x^{12})_\infty
(x^8z^{-2};x^{2r},x^{12})_\infty}{
(x^{12}z^{-2};x^{2r},x^{12})_\infty 
(x^{2r}z^{-2};x^{2r},x^{12})_\infty}\nonumber\\
&\times&
\frac{(x^{2r+-2c+4}z^{-1};x^{2r},x^{12})_\infty
(x^{2c+2}z^{-1};x^{2r},x^{12})_\infty}{
(x^{2r-2c}z^{-1};x^{2r},x^{12})_\infty 
(x^{2c+6}z^{-1};x^{2r},x^{12})_\infty}\nonumber\\
&\times&
\frac{(x^{2r+6-2c+2\pi_{1,2}}z^{-1};x^{2r},x^{12})_\infty
(x^{2c+2\pi_{1,2}}z^{-1};x^{2r},x^{12})_\infty}{
(x^{2r+4-2c-2\pi_{1,2}}z^{-1};x^{2r},x^{12})_\infty 
(x^{2c+2+2\pi_{1,2}}z^{-1};x^{2r},x^{12})_\infty}\nonumber\\
&\times&
\frac{(x^{2r+6-2c+2\pi_{1,3}}z^{-1};x^{2r},x^{12})_\infty
(x^{2c+2\pi_{1,3}}z^{-1};x^{2r},x^{12})_\infty}{
(x^{2r+4-2c-2\pi_{1,3}}z^{-1};x^{2r},x^{12})_\infty 
(x^{2c+2+2\pi_{1,3}}z^{-1};x^{2r},x^{12})_\infty},
\label{def:h}\\
\nonumber
\\
g_1(w)&=&(1-1/w^2)
\frac{(x^{2c+1}/w;x^{2r})_\infty 
(x^{2r-2c-2\pi_{1,2}+1}/w;x^{2r})_\infty}{
(x^{2c+2\pi_{1,2}-1}/w;x^{2r})_\infty 
(x^{2r-2c-1}/w;x^{2r})_\infty
}
,\\
g_2(w)&=&(1-1/w^2)
\frac{(x^{2c+2\pi_{1,2}}/w;x^{2r})_\infty 
(x^{2r-2c+2-2\pi_{1,3}}/w;x^{2r})_\infty}{
(x^{2c+2\pi_{1,3}-2}/w;x^{2r})_\infty 
(x^{2r-2c-2\pi_{1,2}}/w;x^{2r})_\infty}.
\end{eqnarray}
\end{prop}
The parameters $l,k \in P$ are determined by
the boundary conditions.
The parameter $k$ represents the central height.
The parameter $l$ represents the asymptotic boundary
height.
In what follows we consider the case $k=l \in P$ 
for simplicity.
For more general $l,k \in P$,
there exist similar boundary state.
We constract the eigenvector of the commuting
boundary transfer matrix,
$[T_B^{(k)}(z_1),
T_B^{(k)}(z_2)]=0$.
The following is the main result of this paper.
\begin{thm}~~
The boundary state 
$|B\rangle_{k,k}$ is the eigenvector
of boundary transfer matrix $T_B^{(k)}(z)$.
\begin{eqnarray}
T_B^{(k)}(z)|B\rangle_{k,k}=|B\rangle_{k,k}.
\end{eqnarray}
\label{thm:main}
\end{thm}
\begin{cor}~~
Using the type-II vertex operators, we get
general eigenvectors of the boundary transfer matrix $T_B^{(k)}(z)$.
\begin{eqnarray}
&&T_B^{(k)}(z)\cdot \Psi^*_{\mu_1}(\xi_1)\cdots
\Psi^*_{\mu_M}(\xi_M) 
|B\rangle_{k,k}\nonumber\\
&=&\prod_{j=1}^M
\chi(1/z \xi_j)\chi(\xi_j/z)
\cdot
\Psi^*_{\mu_1}(\xi_1)\cdots
\Psi^*_{\mu_M}(\xi_M) 
|B\rangle_{k,k},
\end{eqnarray}
where $\chi(z)$ is given by (\ref{def:chi}).
\end{cor}

Here we sketch proof of main theorem \ref{thm:main}.
Acting the vertex operator $\Phi^{(k+\bar{\epsilon}_j,k)}(z)$
to the condition $T_B^{(k)}|B\rangle_{k,k}=|B\rangle_{k,k}$ from the left,
we get the following necessary and sufficient
condition.
\begin{eqnarray}
z^{\frac{1}{r}(-\frac{r-1}{3}+(\bar{\epsilon}_1|k))} 
h(z)[c-u][\pi_{1,j}+c+u]\Phi_j(z^{-1})|B\rangle_{k,k}=(z
\leftrightarrow z^{-1})~~(j=1,2,3).
\label{ns}
\end{eqnarray}
Here we used the inversion relation of the type-I vertex
operators.
In the case of $j=1$,
after some calculation, LHS becomes the following
\begin{eqnarray}
h(z)h(z^{-1})[c-u][c+u]e^{P(z)+P(1/z)}|B\rangle_{k,k},\nonumber
\end{eqnarray}
which is invariant $z \leftrightarrow z^{-1}$.
Hence the relation (\ref{ns}) for $j=1$ holds.
In the case of $j=2,3$,
after some calculation as similar as
\cite{FK}, 
the relation
(\ref{ns}) are reduced to the following theta identity.
\begin{eqnarray}
&&\frac{[v+k+c-\frac{1}{2}][v-c-\frac{1}{2}]}
{[-v+k+c-\frac{1}{2}][-v-c-\frac{1}{2}]}\\
&=&\frac
{
[c-u][k+c+u][u-v+\frac{1}{2}-k][-v-u+\frac{1}{2}]
-[c+u][k+c-u][-u-v+\frac{1}{2}-k][-v+u+\frac{1}{2}]
}
{[c-u][k+c+u][u+v+\frac{1}{2}-k][v-u+\frac{1}{2}]
-[c+u][k+c-u][-u+v+\frac{1}{2}-k][v+u+\frac{1}{2}]}.\nonumber
\end{eqnarray}

Later the author will write 
complete proof of this theorem,
$U_{q,p}(\widehat{sl}(N,
\mathbb{C}))$ version and generalization of asymptotic
boundary condition $l \in P$
in the another place
\cite{Kojima}.
Our proof is different from those given in \cite{MW}
even for $U_{q,p}(\widehat{sl}(2,{\mathbb C}))$ case.

\section*{Acknowledgement}~
The author would like to thank
the organizing committee of
VIII-th International Workshop Lie theory 
and its applications in Physics, 
held in Varna, Bulgaria 2009.
This work is partly supported by the Grant-in Aid
for Scientific Research {\bf C}(21540228) from Japan Society for
the Promotion of Science.

\begin{appendix}

\section{Some formule}

\begin{eqnarray}
&&h(z)=\nonumber\\
&&\exp\left(
-\sum_{m>0}\frac{1}{2m}\frac{
[(r-1)m]_x[2m]_x}{[rm]_x[3m]_x}z^{-2m}
-
\sum_{m>0}\frac{[(r-1)m]_x}{[rm]_x}
\left(\frac{[2m]_x}{[3m]_x}D_1(m)-\frac{x^{-3m}[m]_x}{
[3m]_x}D_2(m)\right)z^{-m}
\right),\nonumber\\
&&g_1(z)=
\exp\left(
-\sum_{m>0}
\frac{1}{2m}\frac{[(r-1)m]_x (x^m+x^{-m})}{[rm]_x}z^{-2m}
+\sum_{m>0}\frac{[(r-1)m]_x}{[rm]_x}(D_1(m)-x^{-2m}
D_2(m))z^{-m}\right),
\nonumber
\\
&&g_2(z)=\exp\left(
-\sum_{m>0}
\frac{1}{2m}\frac{[(r-1)m]_x (x^m+x^{-m})}{[rm]_x}z^{-2m}
+\sum_{m>0}\frac{[(r-1)m]_x}{[rm]_x}D_2(m)x^{-m}z^{-m}\right).
\nonumber
\end{eqnarray}

~\\

\end{appendix}


\begin{thebibliography}{99}
\bibitem{DFJMN}B.Davis, O.Foda, M.Jimbo, T.Miwa and
A.Nakayashiki,
Diagonalization of the XXZ Hamiltonian
by vertex operators,
{\it Commun.Math.Phys.} 
{\bf 151}, 89- (1993).
\bibitem{JM}M.Jimbo and T.Miwa,
Algebraic Analysis of Solvable Lattice Models,
CBMS Regional Conference Series in Mathematics,
Vol {\bf 85}, Providence, RI:AMS, (1994).
\bibitem{Sklyanin}E.K.Sklyanin,
Boundary condition for integrable quantum system,
{\it J.Phys.}{\bf A21}, 2375- (1988).
\bibitem{ABF}G.E.Andrews, R.J.Baxter and P.J.Forerester,
Eight-vertex SOS model and generalized
Rogers-Ramanujan-type identities,
{\it J.Stat.Phys.}{\bf 35}, 193- (1984).
\bibitem{JMO}M.Jimbo, T.Miwa and M.Okado,
Local state probabilities of solvable lattice models
: an $A_{n-1}^{(1)}$ family,
{\it Nucl.Phys.}{\bf B300} [FS22], 74- (1988).
\bibitem{JKKKM}M.Jimbo,R.Kedem,T.Kojima,H.Konno and T.Miwa,
XXZ chain with a boundary,
{\it Nucl.Phys.}{\bf B441}[FS], 437- (1995).
\bibitem{MW}T.Miwa and R.Weston,
Boundary ABF models,
{\it Nucl.Phys.}{\bf B486}[PM], 517- (1997).
\bibitem{BFKZ}M.T.Batchelor, V.Fridkin, A.Kuniba, Y.K.Zhou,
Solutions of the reflection equation for face and vertex models
associated with $A_n^{(1)}, B_n^{(1)}, C_n^{(1)}$ and $A_n^{(2)}$,
{\it Phys.Lett.}{\bf B376} no.4, 266- (1996).
\bibitem{FK}H.Furutsu and T.Kojima,
The $U_q(\widehat{sl_n})$ analogue of the XXZ chain with a boundary,
{\it J.Math.Phys.}
{\bf 41}, no.7,
4413- (2000).
\bibitem{LP}S.Lukyanov and Ya.Pugai,
Multi-point local height probabilities in the integrable RSOS
model,{\it Nucl.Phys.}{\bf B473}[FS], 631- (1996).
\bibitem{AJMP}Y.Asai, M.Jimbo, T.Miwa, Ya.Pugai,
Bosonization of vertex operators for the $A_{n-1}^{(1)}$
face model,
{\it J.Phys.}{\bf A29}, 6595- (1996).
\bibitem{FKQ}H.Furutsu, T.Kojima and Y.Quano,
Type-II vertex operators for the $A_{n-1}^{(1)}$ face model,
{\it Int.J.Mod.Phys.}{\bf A15},no.10, 1533- (2000).
\bibitem{JKOS}M.Jimbo, H.Konno, S.Odake and J.Shiraishi :
Elliptic Algebra $U_{q,p}(\widehat{sl_2})$ : Drinfeld currents and vertex operators,
{\it Commun.Math.Phys.}{\bf 199}, 605- (1999).
\bibitem{KK}T.Kojima and H.Konno :
The Elliptic Algebra $U_{q,p}(\widehat{sl_N})$ and the Drinfeld Realization of the Elliptic Quantum Group
${\cal B}_{q,\lambda}(\widehat{sl_N})$,
{\it Commun.Math.Phys.}{\bf 237}, 405- (2003).
\bibitem{Kojima}T.Kojima , in preparation.
\end{thebibliography}
\end{document}